\documentclass[review]{elsarticle}

\usepackage{lineno,hyperref}
\usepackage{graphicx}
\usepackage{array}
\usepackage{amsmath}
\modulolinenumbers[5]

\journal{Optics Communications}

\begin{document}

\begin{frontmatter}

\title{Enhancing the photon-extraction efficiency of site-controlled quantum dots by deterministically fabricated microlenses}


\cortext[cor]{Corresponding author}
\ead{stephan.reitzenstein@tu-berlin.de}
\author[mymainaddress]{Arsenty Kaganskiy}
\author[mymainaddress]{Sarah Fischbach}
\author[mymainaddress,mysecondaryaddress]{Andr\'{e} Strittmatter}
\author[mymainaddress]{Sven Rodt}
\author[mymainaddress]{Tobias Heindel}
\author[mymainaddress]{Stephan Reitzenstein\corref{cor}}

\address[mymainaddress]{Institut f{\"u}r Festk{\"o}rperphysik, Technische Universit{\"a}t Berlin, Hardenbergstra{\ss}e 36, D-10623 Berlin, Germany}
\address[mysecondaryaddress]{Present address: Abteilung Halbleiterepitaxie, Otto-von-Guericke Universit{\"a}t Magdeburg,  Universit{\"a}tsplatz 2, D-39106 Magdeburg, Germany}

\begin{abstract}
We report on the realization of scalable single-photon sources (SPSs) based on single site-controlled quantum dots (SCQDs) and deterministically fabricated microlenses. The fabrication process comprises the buried-stressor growth technique complemented with low-temperature in-situ electron-beam lithography for the integration of SCQDs into microlens structures with high yield and high alignment accuracy. The microlens-approach leads to a broadband enhancement of the photon-extraction efficiency of up to (21~$\pm$~2)~$\%$ and a high suppression of multi-photon events with g$^{(2)}$($\tau$ = 0) $<$ 0.06 without background subtraction. The demonstrated combination of site-controlled growth of QDs and in-situ electron-beam lithography is relevant for arrays of efficient SPSs which can be applied in photonic quantum circuits and advanced quantum computation schemes.  
\end{abstract}


\end{frontmatter}


\section{Introduction}

The development and optimization of quantum light sources is a central topic of quantum nanophotonics and advanced quantum communication. The latter requires close to ideal properties in terms of suppression of multi-photon emission, photon indistinguishability, and entanglement fidelity to implement for instance the quantum repeater protocol for long-distance quantum communication \citep{Aharonovich.2016}. In principle, self-assembled quantum dots (QDs) are supreme candidates to meet these stringent requirements. However, due to their random positions and emission energies it is a great technological challenge to integrate single QDs in a well-controlled way into photonic structures to enhance their photon-extraction efficiency and to improve their quantum nature of emission. Popular approaches for efficient light outcoupling include photonic wires \citep{Claudon.2010}, micropillar  cavities\citep{Ding.2016} and microlenses \citep{Gschrey.2015}. In parallel, during the last decade deterministic nano$\-$processing technologies based on optical and electron-beam in-situ lithography have been developed and refined in order to integrate single self-assembled QDs into high-quality micropillar cavities\citep{Somaschi.2016}, circular dielectric gratings\citep{Sapienza.2015} and microlenses\citep{Gschrey.2015}. Although this deterministic nano$\-$technology approach has been exploited very successfully for the realization of sources generating indistinguishable single-\citep{Thoma.2016} or twin-\citep{Heindel.2017} photon states based on self-organized QDs, it is less suitable for the realization of ordered arrays of quantum light sources, due to the random spatial position of the selected QDs. This represents a major hurdle for the realization of parallel optical links for, e.g., photonic quantum circuits that require a multitude of matching single-photon sources (SPSs). A prominent example is boson sampling on a photonic chip based on nonclassical interference of photons in an integrated photonic circuit, \citep{Spring.2013} and quantum computing based on photonic coupling in an array of QDs.\citep{Jones.2012} In order to upscale the realization of QD-based SPSs and to eventually enable such appealing applications, one needs to apply advanced growth techniques which allow for the site-controlled nucleation of single QDs. Prominent examples for the realization of site-controlled QDs are based on top-down etching techniques,\citep{Demory.2015} site-selective growth on nanohole-arrays, \citep{Schneider.2009, Baumann.2012, Jons.2013} inverted pyramids \citep{Pelucchi.2007, Surrente.2009} and buried-stressors\citep{Strittmatter.2012} respectively. In this Letter we present a concept which combines the advantages of site-controlled growth, in-situ electron-beam lithography and microlenses to realize highly efficient single-photon emitters in a potentially scalable nano$\-$technology platform.   

We apply the buried stressor approach which is based on the selective oxidation of an AlAs add-layer in a semiconductor heterostructure.\citep{Strittmatter.2012b} The controlled oxidation allows for stress engineering in the overlying layers \citep{Kieling.2015} due to a reduced volume of the oxidized layer by about 12-13 $\%$.\citep{Takamori.1996} During a subsequent overgrowth, In(Ga)As QDs tend to nucleate at the maximum tensile strain emerging over the underlying AlAs aperture. Hereby, the diameter of the aperture influences the strain distribution at the surface in a way that for small aperture diameters and proper adjusted QD growth parameters single QDs can be grown above the oxide aperture. Compared to QDs aligned to etched nanohole-arrays, this type of site-controlled QDs (SCQDs) does not suffer from close-by etched surfaces, enabling a high optical quality similar to that of standard self-assembled QDs.\citep{Strau.2017} This advantage is obtained at the cost of a rather gentle site control which can lead to a slight displacement from the aperture's center towards the aperture's boundaries between the oxidized and non-oxidized regions. Therefore, the precise determination of the SCQD's position is inevitable before fabricating a microlens since a lateral displacement $\geq$ 50~nm of the emitter would lead to a significant decrease of the extraction efficiency.\citep{Gschrey.2015} In-situ EBL allows for an alignment accuracy of 34~nm \citep{Gschrey.2015b} and, hence, fulfills the aforementioned requirements.

\section{Sample fabrication and simulations}

\begin{figure}[t!]
\centering
\includegraphics[width=11 cm]{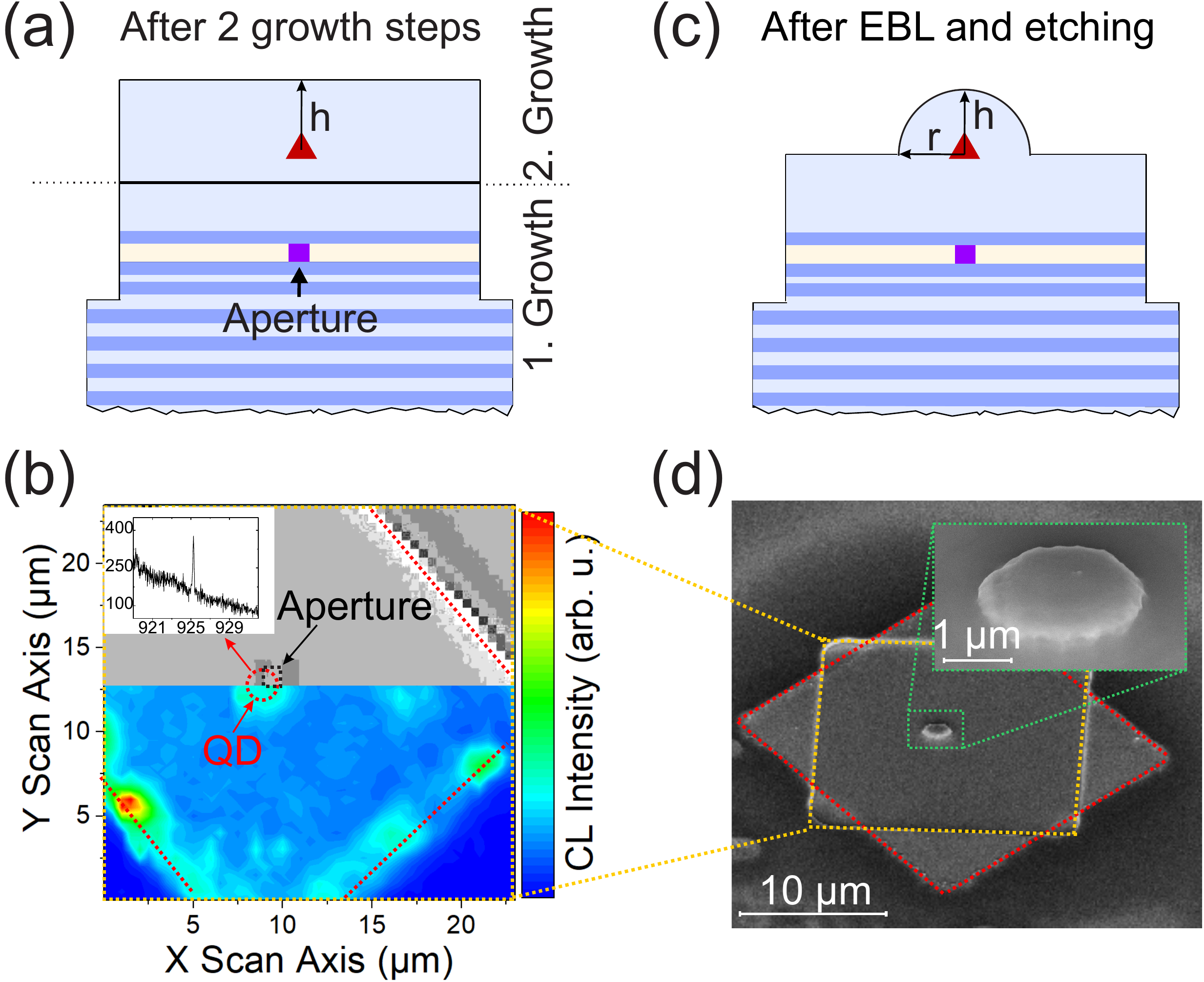}
\caption{Two excerpts of the work flow for fabricating SCQDs with deterministically placed microlenses: The two-step growth of the epitaxial structure via the buried stressor approach (a) is followed by microlens processing via in-situ EBL and subsequent ICP-RIE dry etching (c). (b) SEM image of a structure as shown in (a) with a recognizable aperture (top part) and 2D cathodoluminescence map of the same structure with the pronounced luminescence from the SCQD (bottom part) as described in the text (positions of the QD and the aperture are marked with a dashed circle and square, respectively). The spectrum of the corresponding QD taken during the in-situ EBL process is shown in the inset. (d) SEM image of a fully processed microlens structure (cf. inset). The mesa whose cross-section is sketched in (a) and (c) and the CL mapping area are marked in (b) and (d) with a dashed red and yellow rectangle, respectively.}
\label{fig:fabrication}
\end{figure}$•$

\begin{figure}[t!]
\centering
\includegraphics[width=12 cm]{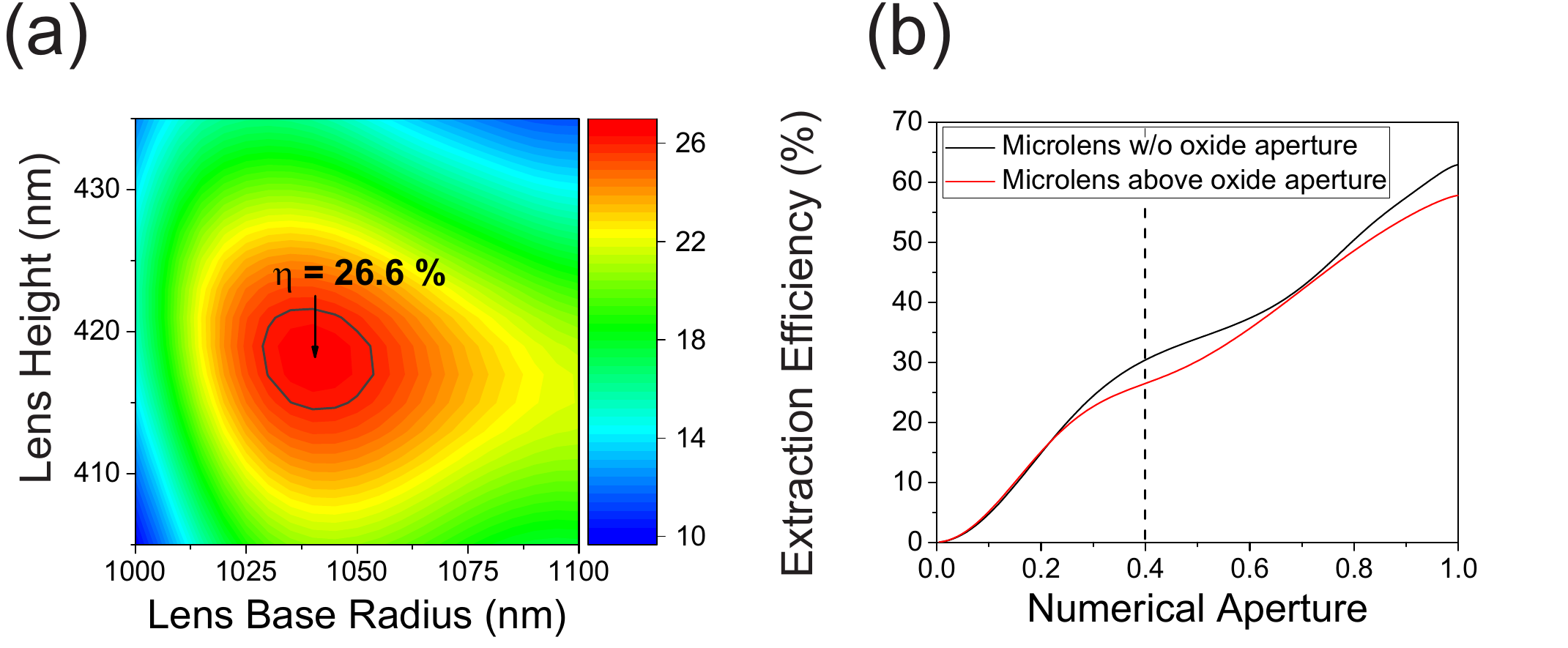}
\caption{Numerical optimization of the microlens parameters to maximize the light-extraction efficiency $\eta$: (a) 2D map of $\eta$ as a function of the lens radius and height . The highest extraction efficiency of 26.6 $\%$ as well as the region with $\eta$ $>$ 26 $\%$ are marked. (b) Photon-extraction efficiency $\eta$ as a function of the numerical aperture (NA) of the collection optics for a standard microlens structure and a microlens above an oxide aperture.}
\label{fig:simulations}
\end{figure}

The sample fabrication is schematically presented in Fig.~\ref{fig:fabrication}~(a) and (c). It starts with the growth of the Al(Ga)As heterostructure via metalorganic chemical vapor deposition (MOCVD) on an n-doped GaAs (100) substrate based on the numerically optimized layer design (details on the simulation will be given below). First, a 300 nm thick GaAs buffer layer and the DBR consisting of 27 pairs of $\lambda$/4 thick $\textrm{Al}_{\textrm{0.90}}\textrm{Ga}_{\textrm{0.10}}\textrm{As}$/GaAs are grown followed by a 30 nm thick AlAs layer sandwiched in 40 nm thick $\textrm{Al}_{\textrm{0.90}}\textrm{Ga}_{\textrm{0.10}}\textrm{As}$ claddings. The first growth step is finalized by an 80~nm thick GaAs capping layer. Next, an array of quadratic mesa structures with a base width of 20 to 21 $\mu$m is processed with a pitch of 260 $\mu$m into the template by UV lithography and inductively-coupled plasma reactive-ion dry etching (ICP-RIE). After selective oxidation of the apertures the second epitaxial growth step is performed. It starts with the growth of a 50 nm thick GaAs layer followed by the Stranski-Krastanow growth of InAs \linebreak SCQDs and a 418 nm thick GaAs capping layer which corresponds to the numerically optimized height of the lenses (cf. Fig. \ref{fig:fabrication} (a)). Subsequently, the sample is spin-coated with 110 nm of the electron-beam resist CSAR 62\citep{Kaganskiy.2016} followed by in-situ EBL.\citep{Gschrey.2015,Kaganskiy.2015} The bottom part of Fig. \ref{fig:fabrication} (b) shows exemplarily a 2D cathodoluminescence (CL) map of the mesa surface taken during the CL lithography step. The map refers to the spectral range of 925.21 to 925.49 nm and indicates the pronounced luminescence of the charged X-line of the SCQD whose spectrum is shown in the inset. The position of this SCQD is marked by a dashed red circle in Fig.~\ref{fig:fabrication}~(b). At the edges of the overgrown mesa emission spots of defect centers are visible. At the same time a scanning electron microscope (SEM) image is recorded (top part of the figure) demonstrating the aperture which can be identified due to a surface modulation caused by the modified strain in the underlying layers. A comparison shows the moderate misalignment between the center of the aperture and the SCQD of about~800 nm. During in-situ EBL, lenses with the optimized height (h~=~418~nm) and radius (r~=~1040~nm) (cf. Fig.~\ref{fig:simulations}~(a)) are written into the resist at the positions of the identified SCQDs. The CL mapping is performed with an exposure dose corresponding to the positive-tone regime of the used resist while the lens writing is done in the negative-tone regime\citep{Kaganskiy.2016} resulting in a removal of the resist from the scanned area except for the lens position in the following development step (the mapped area is marked by a yellow dashed rectangle in Fig.~\ref{fig:fabrication}~(b) and (d)). The fabrication is finalized by ICP-RIE (cf. Fig.~\ref{fig:fabrication}~(c)) followed by an SEM-based analysis of the lateral displacement of the microlenses  with respect to the aperture's centers. Fig.~\ref{fig:fabrication}~(d) presents an SEM image of the fully processed device and a zoom-in of the microlens (inset). It is noteworthy, that the presented approach provides a high yield for the realization of SCQDs. In the present study 17 out of 27 mesa structures contained a SCQD corresponding to a yield of 63~$\%$. The average displacement of the SCQDs from the center of the oxidized aperture was determined to be 640~nm for a given aperture side-length in the range between 1200 and 1600 nm.

Prior to the above-described sample fabrication, the structural design was optimized in order to maximize the photon-extraction efficiency $\eta$. The simulation results presented in Fig. \ref{fig:simulations} were obtained using the finite-element software package JCMsuite.\citep{JCMwaveGmbH.} In the simulations we consider an oxide aperture of 800~nm side-length which is typical for the processing of SCQDs. The aperture is integrated in the uppermost DBR pair while the distance of $\lambda$/2n, where $n$ is the refractive index of GaAs, between the aperture and the dipole emitter allows for the dipole alignment with the antinode of the electrical field in vertical direction. The height and radius of the microlens are varied in the simulations to maximize $\eta$ while the emitter is located in the center of the lens base. Fig.~\ref{fig:simulations}~(a) shows the result of a corresponding parameter scan performed for an emission wavelength of 925~nm and a numerical aperture NA~=~0.4 of the collection optics. We find a rather broad region with an extraction efficiency $\eta$~$\textgreater$~26~$\%$ and a maximum of 26.6~$\%$ for a lens height of 418~nm and a lens radius of 1040~nm. Additionally, Fig.~\ref{fig:simulations}~(b) displays $\eta$ as a function of the NA. It is noteworthy, that the lowering of the extraction efficiency in case of a structure with a microlens above an oxide aperture as compared to a standard microlens structure is very moderate and becomes negligible at higher numerical apertures around 0.7 which could be tackled experimentally by the on-chip integration of high-NA micro-objectives.\citep{Fischbach.2017b, Gissibl.2016}

\section{Results and discussion}

The microlens structures are investigated via micro-photoluminescence ($\mu$PL) spectroscopy measurements to determine the photon-extraction efficiency and the quantum nature of emission. The sample is mounted inside a He-flow cryostat at a temperature of 7.5~K. Optical excitation is performed by a mode-locked Titanium:sapphire laser emitting at a wavelength of 808 nm (pulse width: 2~ps, repetition frequency: 80~MHz). The resulting photoluminescence is collected via a microscope objective with an NA of 0.4, spectrally dispersed by a monochromator and finally detected by a Silicon-based charge-coupled device camera, with an overall spectral resolution of 30 $\mu$eV. Photon autocorrelation measurements are performed by means of a fiber-based Hanbury-Brown and Twiss (HBT) setup with single-photon counting modules based on silicon avalanche photodiodes (temporal resolution: 350~ps).  

\begin{figure}[t!]
\centering
\includegraphics[width=11 cm]{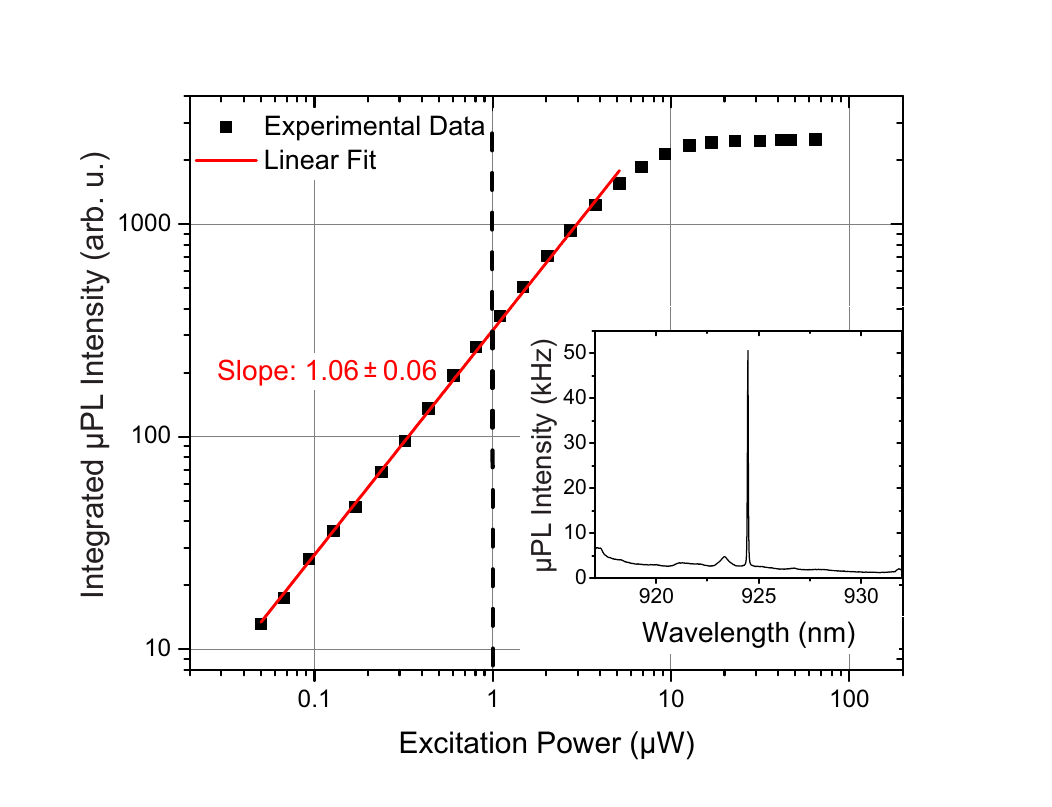}
\caption{$\mu$PL investigations on a SCQD integrated in a microlens: integrated $\mu$PL intensity in logarithmic scaling as a function of the excitation power $P$. The linear rise of 1.06~$\pm$~0.06 is typical for the emission of a single exciton. The dashed line marks the excitation power at which the autocorrelation and the time-resolved measurements were performed. Inset: spectrum of the corresponding QD line taken at $P$ = 16.9 $\mu$W.}
\label{fig:spec-power}
\end{figure}

The inset of Fig. \ref{fig:spec-power} shows a $\mu$PL spectrum detected at an excitation power of 16.9 $\mu$W. The single emission line with a resolution limited linewidth is associated with a trion state of the QD presented above (cf. Fig. \ref{fig:fabrication} (b)) in the center of the microlens. The corresponding excitation-power-dependent integrated intensity of this excitonic transition is plotted in the main panel of Fig. \ref{fig:spec-power} on a double logarithmic scale.  The power-scaling of this line with a slope of 1.06~$\pm$~0.06 together with the absence of the fine structure splitting extracted from a polarization-resolved measurement (not shown here) supports the interpretation of single excitonic emission of the charged trion state.  Taking into account a setup efficiency of 0.8 $\%$ we determined a photon-extraction efficiency of $\eta$~=~(21~$\pm$~2)~$\%$ for the investigated QD at saturation pump power (both numbers were measured and calculated analog to the procedure described in Ref. \citep{Gschrey.2015}). A moderate deviation of the determined extraction efficiency from the theoretically calculated value of 26.6~$\%$ can be attributed to aberrations in radius and height of the lens during the dry-etching step as reported previously.\citep{Fischbach.2017} This record-high photon-extraction efficiency for SPSs based on SCQDs reflects a high quality of the structures and the fabrication accuracy achieved by in-situ EBL.

\begin{figure}[t!!]
\centering
\includegraphics[width=11 cm]{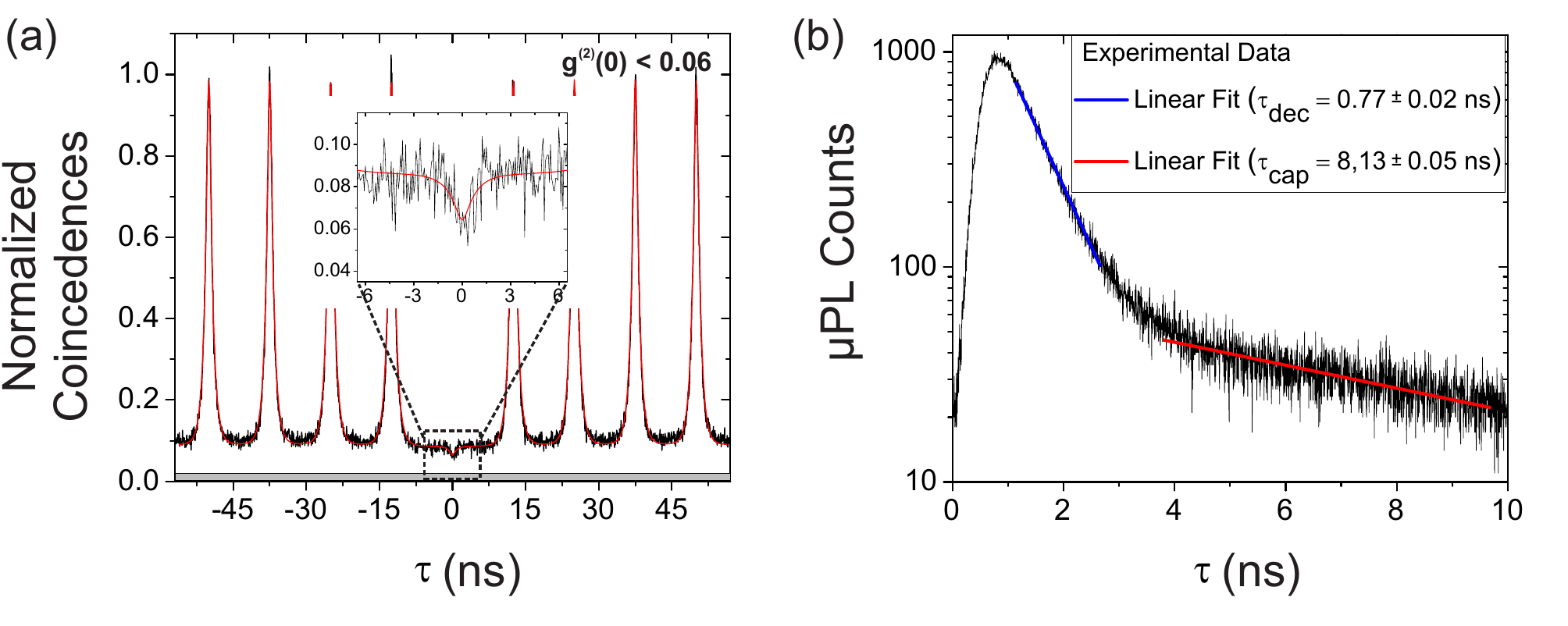}
\caption{(a) Photon autocorrelation measurement performed under pulsed laser excitation at 1 $\mu$W indicating a dip with g$^{(2)}$($\tau$ = 0) $<$ 0.06. After background subtraction (marked by the grey area), the fitting of the experimental data according to Eq. \ref{equation1} yields g$^{(2)}$($\tau$ = 0) = 0.03 $\pm$ 0.01. The inset shows a zoom-in of the area at $\tau$ = 0 showing a characteristic behavior for recapture processes. (b) Time-resolved measurement showing a biexponential behavior with time decay constants $\tau_{dec}$ = (0.77 $\pm$ 0.02) ns and $\tau_{cap}$ = (8.13 $\pm$ 0.05) ns.} 
\label{fig:g2-time}
\end{figure}

The quantum nature of emission of the SCQD-microlens is probed by performing a photon-autocorrelation measurement at an excitation power of 1 $\mu$W. Fig. \ref{fig:g2-time} (a) shows the autocorrelation histogram indicating a dip at zero time delay with g$^{(2)}$($\tau$ = 0) $<$ 0.06 (cf. inset of Fig. \ref{fig:g2-time} (a)). The observed antibunching is limited by recapture processes which are typical for non-resonant excitation as applied here and that lead to a repopulation of the QD with charge carriers with a small but finite probability and with a time delay of $\tau_{cap}$.\citep{Peter.2007,Kumano.2016} Then, after the decay time $\tau_{dec}$ the secondary photon emission occurs resulting in the aforementioned maxima at small $\tau$. The g$^{(2)}$($\tau$) at small time delays can be described with the following expression:\citep{Dalgarno.2008, Fischbach.2017b}
\begin{equation}
g^{(2)}(\tau) \propto \exp{(\frac{-|\tau|}{\tau_{dec}}}) - \exp{(\frac{-|\tau|}{\tau_{cap}}}) + B
\label{equation1}
\end{equation}
with the offset B determining the background (shown as a grey area in Fig. \ref{fig:g2-time} (a)) taking into account the APDs' dark counts and the laser background.\citep{Brida.2011} The aforementioned time constants are determined independently by means of a time-resolved measurement (Fig. \ref{fig:g2-time} (b)) resulting in $\tau_{dec}$ = (0.77 $\pm$ 0.02) ns and $\tau_{cap}$ = (8.13 $\pm$ 0.05) ns. Finally, the measured data is fitted with a function based on Eqn. (\ref{equation1}) for the central region and a biexponential function including the time constants $\tau_{dec}$ and $\tau_{cap}$ for the side peaks (red line in Fig. \ref{fig:g2-time} (a)) resulting in g$^{(2)}$($\tau$ = 0) = 0.03 $\pm$ 0.01 with B = 0.02.

\section{Conclusions}

In conclusion we presented a novel fabrication platform which allows for the deterministic integration of SCQDs into microlenses. This scheme is highly attractive since it leads to an enhanced photon-extraction efficiency without the need for a complex overgrowth of the SCQDs with a DBR which would be required in case of cavity enhanced SPSs.\citep{Schneider.2009} The fabricated devices show a record-high extraction efficiency for SPSs based on SCQDs of up to (21~$\pm$~2)~$\%$ into collection optics with an NA of 0.4 as well as a very high suppression of multi-photon emission events with g$^{(2)}$($\tau$ = 0) = 0.03 $\pm$ 0.01 ($<$ 0.06 without background subtraction). The presented approach of combining SCQDs with in-situ EBL can be readily applied for other types of SCQDs based on nanohole arrays or inverted pyramids to enhance their extraction efficiency, which could pave the way for scalable and regular arrays of SPSs to enable photonic quantum circuits and quantum computation based on optically coupled QDs.

\section*{Acknowledgments}

The research leading to these results received funding from the European Research Council under the European Union\textsc{\char13}s Seventh Framework Program Grant Agreement No. 615613, from the Volkswagen Foundation via NeuroQNet and 
from the German Research Foundation via CRC 787.

\end{document}